# Heat guiding and focusing using ballistic phonon transport in phononic nanostructures

Roman Anufriev[*,1], Aymeric Ramiere[1,2], Jeremie Maire[1], and Masahiro Nomura[*,1,3]

[1] Institute of Industrial Science, the University of Tokyo, Tokyo, 153–8505, Japan
[2] Laboratory for Integrated Micro Mechatronic Systems / National Center for Scientific Research-Institute of Industrial Science (LIMMS/CNRS-IIS), the University of Tokyo, Tokyo, 153–8505, Japan
[3] PRESTO, Japan Science and Technology Agency, Saitama, 332–0012, Japan

[*]Email: anufrievroman@yandex.com, nomura@iis.u-tokyo.ac.jp

**Abstract**

*Unlike classical heat diffusion at the macroscale, nanoscale heat transport can occur without energy dissipation because phonons can travel in straight lines for hundreds of nanometres. Despite recent experimental evidence of such ballistic phonon transport, control over its directionality, and thus its practical use, remains a challenge, as the directions of individual phonons are chaotic. Here, we show a way to control the directionality of ballistic phonon transport using silicon thin-films with arrays of holes. First, we demonstrate the formation of directional heat fluxes in the passages between the holes. Next, we use these nanostructures as a directional source of ballistic phonons and couple the emitted phonons into nanowires. Finally, we introduce a nanoscale thermal lens in which the phonons converge at a focal point, thus focusing heat into a spot of a few hundred nanometres. These results provide a basis for ray-like heat manipulations that enable nanoscale heat guiding, dissipation, localization, confinement and rectification.*

Studies of nanoscale heat transport in semiconductors are largely motivated by overall miniaturization of microelectronic devices[1] and the search for particular nanostructures suitable for thermoelectrics[2], heat localization[3], thermal rectification[1,4] and other applications[5]. At the nanoscale, however, heat transport can no longer be described by classical diffusion along temperature gradients because individual phonons can travel in straight lines without energy dissipation for hundreds of nanometres. Such point-to-point propagation of phonons between diffuse scattering events is known as *ballistic* phonon transport[6–8] and has recently been detected[9–14] in various nanostructures and materials.

Yet, its practical use remains challenging, as different phonons travel in different directions, making the overall heat flux still simply following the temperature gradient. Once control over directionality is achieved, the possibility to guide and locally apply heat without dissipation can be used in biomedicine[15–18], thermoelectrics[19], phase-change material technology[20,21], chemical reaction control[22] and virtually wherever wireless nano-heaters are required.

The solution may lie in the novel class of thermal materials called phononic crystals[23–25]—nanostructures typically consisting of thin-films with two-dimensional array of holes that can be designed to control propagation of phonons. Room temperature heat transport in phononic crystals is mostly dominated by diffuse surface scattering of phonons[26–28], yet some phonons can travel ballistically over hundreds of nanometres in cross-plane direction[9]. In this work, we aim to shape the paths of these ballistic phonons via nano-patterning in order to control in-plane ballistic heat transport. Using micro time-domain thermoreflectance (μ-TDTR) experiments and Monte-Carlo simulations, we show the formation of directional fluxes of ballistic phonons in silicon phononic crystals and use this effect for directional heat guiding and focusing at the scale of a few hundred nanometres.

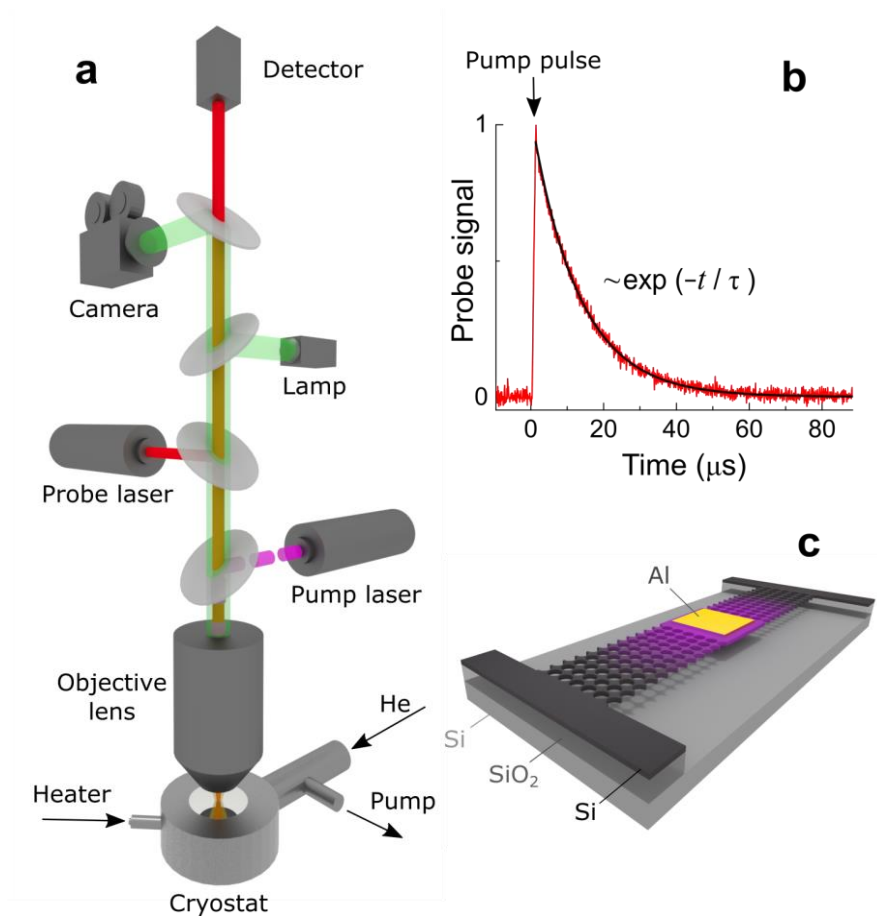

**Figure 1. Experimental setup and a typical sample for thermal decay measurements**. **a**, The pump and probe beams are focused on the sample by an objective lens; the detector monitors changes in the reflection of the probe beam (ΔR/R) in time, caused by the pulse of the pump beam. **b**, The signal is fitted by an exponential function. **c**, Schematic of a typical air-bridged sample.

## Experimental technique

To study nanoscale heat transport, we used an originally developed μ-TDTR technique[29,30] (Methods). This all-optical experimental method is a powerful tool for contactless thermal measurements on a large number of samples. Figure 1 shows schematics of our μ-TDTR setup and a typical sample. The samples are mounted in a high-vacuum helium-flow cryostat that enables a precise temperature control. A microscope objective lens focuses two laser beams on an aluminium pad on top of each sample; the pulsed *pump* beam periodically heats the aluminium pad, while the change in its reflectance (ΔR/R), caused by the heating, is monitored by the continuous-wave *probe* beam. Since the change in reflectance is proportional to the change in temperature via the thermoreflectance coefficient, we can record relative changes in the temperature (ΔT/T) of the aluminium pad in time ($t$). Figure 1b shows that a measured thermal decay curve can always be well fitted by an exponential decay *exp* $(-t/\tau)$, thus the only quantity that characterizes each sample is the thermal decay time ($\tau$) – the time for heat to dissipate from the aluminium pad through the sample. To ensure dissipation through the structure of interest only, the investigated structures are suspended, as shown in Figure 1c.

To create such structures, we used a standard top-down approach on a silicon-on-insulator wafer with a 145-nm-thick top layer (Methods). First, we deposited $4 \times 4$ μm$^2$ aluminium pads in the centres of the future structures. Next, the phononic structures were formed by electron beam lithography followed by reactive ion etching. Finally, the underlying SiO$_2$ layer was removed using vapour etching, creating 5-μm-wide and 145-nm-thick air-bridged structures, with the length depending on the type of the structure: 25 μm for aligned/staggered samples; 12 μm plus nanowire length for nanowire-coupled samples; 16 μm for thermal lenses.

### Heat transport in aligned and staggered lattices

First, we investigated the possibility of in-plane ballistic heat transport in phononic crystals. One way to detect the presence of ballistic phonons is to compare structures in which phonons can freely propagate with those in which direct passage is blocked. Thus, we studied two types of hole arrangement: an aligned (square) lattice and a staggered lattice, in which every second row of holes is shifted by a half-period from its position in the aligned lattice. To access heat transport at different scales, we fabricated samples with periods of 160, 200, 280, 350 and 500 nm, shown in Figures 2a-d, and several diameter-to-period ($d/a$) ratios for each period.

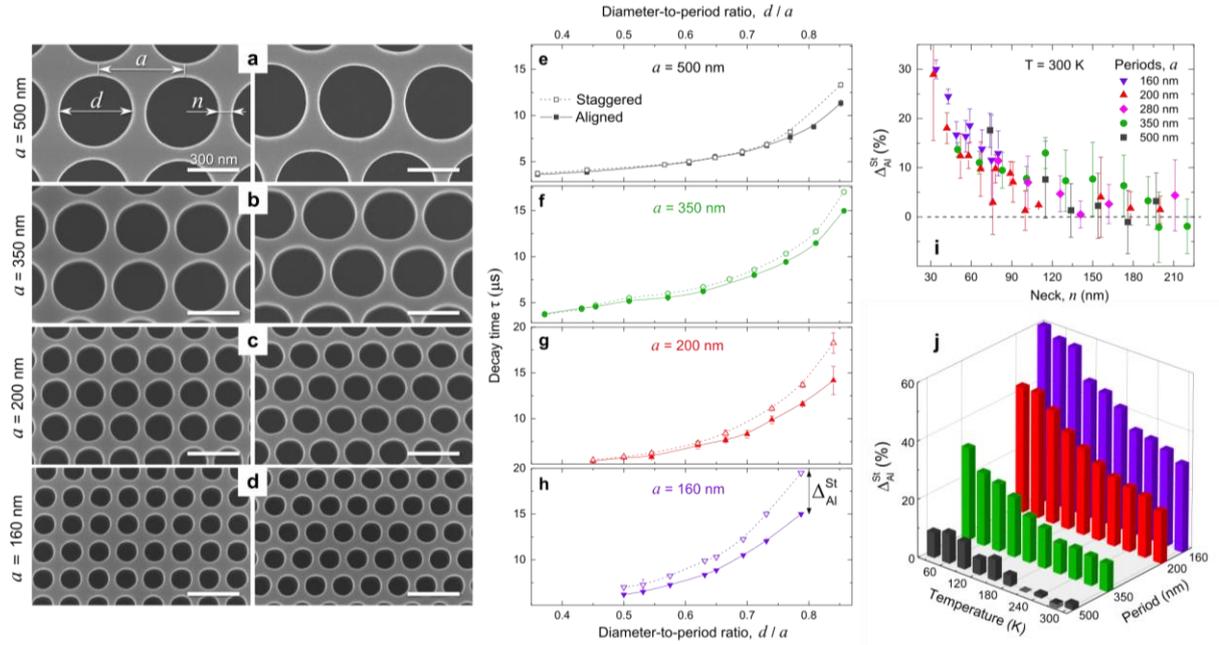

**Figure 2. Thermal conduction in aligned and staggered lattices**. **a-d**, SEM images of typical aligned and staggered lattice samples with periods of 500, 350, 200 and 160 nm. Scale bars are 300 nm. **f-d**, Decay times measured at room temperature in aligned samples deviate from those in staggered ones as the period is changed from 500 to 350, 200 and 160 nm and as the diameter-to-period ratio is increased. Lines are guides for the eye. **i**, Difference between decay times of aligned and staggered samples ($\Delta_{Al}^{St} = (\tau_{St} - \tau_{Al})/\tau_{Al}$) correlates with the neck ($n$), and **j**, strengthens by a factor of two as the temperature is decreased to 30 K. The error bars depict the standard deviation in multiple measurements (Methods) and are less than the size of the point for most of the data in **e-h** and less than ± 7% in panel j.

At the microscale ($a$ = 500 nm), the thermal decay times ($\tau$) measured on the samples with aligned and staggered lattices were indistinguishable (Fig. 2e), at least as long as the diameter-to-period ratio was below 0.75. This result is consistent with the pioneering work by Song and Chen[31] on aligned and staggered microporous structures ($a$ = 4 μm and $d/a$ = 0.57), and confirms that microscale heat transport is mostly diffusive[32]. Yet, even in these large samples, the decay times in the aligned lattice deviated from those in the staggered lattice for $d/a > 0.75$. This deviation was more and more pronounced as the period was reduced to 350 nm (Fig. 2f) and further to 200 nm (Fig. 2g). Finally, as we scaled the structures down to 160 nm in period, the difference between the aligned and staggered lattices became evident for all measured samples (Fig. 2h). In other words, at the nanoscale, heat dissipated faster through the aligned than through the staggered lattice.

Remarkably, the difference between aligned and staggered lattices did not simply appear at a certain porosity, regardless of the scale, but seemed to depend on some characteristic dimension of the structure. One such dimension, impacting heat transfer in phononic crystals[29,33], is the distance between two adjacent holes, called the *neck* (*n*). Figure 2i shows that the difference between decay times in the aligned ($\tau_{Al}$) and staggered ($\tau_{St}$) samples $\Delta_{Al}^{St} = (\tau_{St} - \tau_{Al})/\tau_{Al}$ seems to correlate with the neck size: regardless of the period, a clear difference appears only when the neck becomes smaller than 100 nm and increases up to 30% as the neck is reduced to 30 nm. In terms of thermal conductivity, its value in the smallest samples (*a* = 160 nm, *d* / *a* = 0.79) is reduced from 20.1 Wm$^{-1}$K$^{-1}$ in the aligned lattice to 16.5 Wm$^{-1}$K$^{-1}$ in the staggered lattice (Supplementary Information). Thus, staggering is an effective way to reduce the thermal conductivity, which can further improve performance of thermoelectric devices based on phononic nanostructures[34].

Next, we repeated the measurements at different temperatures in the 30 – 300 K range on four characteristic pairs of samples, one of each period (*d* / *a* ≈ 0.73 for *a* = 500 and 350 nm; *d* / *a* ≈ 0.79 for *a* = 200 and 160 nm). We found that the effect was temperature dependent and strengthened by a factor of two as the temperature was decreased from 300 to 30 K (Fig. 4j). At 30 K the difference $\Delta_{Al}^{St}$ eventually appeared even in the largest samples, while in the smallest ones it became as high as 60%.

### Ballistic heat transport in phononic crystals

These observations cannot be explained by classical heat diffusion (Supplementary Information), thus we assume that at the nanoscale heat no longer spreads diffusively, but propagates at least partly ballistically. Indeed, in the aligned lattice, phonons could in principle travel ballistically in the passages between the holes, whereas in the staggered lattice these passages are reduced or even blocked completely.

To verify this hypothesis, we simulated the propagation of phonons in our samples ($a$ = 350 nm, $d/a$ = 0.85, $\eta$ = 2 nm) at 4 K using a specially developed two-dimensional Monte-Carlo technique (Methods). Figure 3 shows the thermal energy distribution in aligned and staggered structures and corresponding spatial and angular distributions of phonons after ten rows of holes. In the staggered lattice, phonons are scattered by the staggered holes; this scattering reveals itself as faint yellow regions just under the upper row of holes (Fig. 3a). Consequently, at the end of the structure phonons are uniformly scattered in space and have a broad range of exit angles (Fig. 3b).

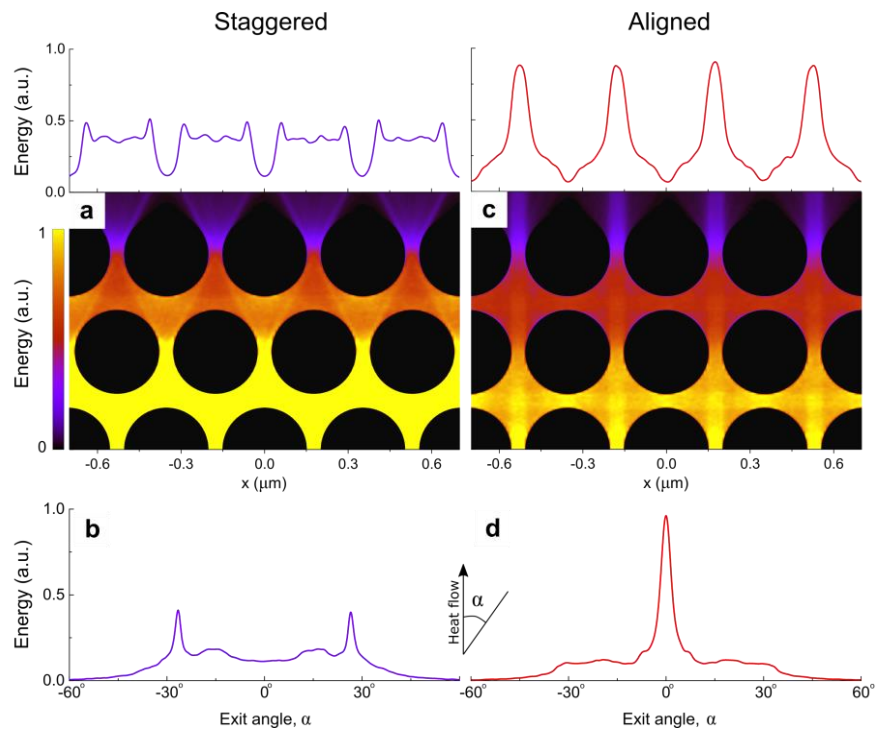

**Figure 3**. **Heat flow directionality in aligned and staggered lattices**. Monte-Carlo simulations predict rather uniform spatial and angular energy distributions in the staggered lattice **a-b**, but strong anisotropy in the aligned lattice **c-d**.

Conversely, in the aligned lattice, phonons develop a certain directionality and fluxes of heat appear in the passages between holes (Fig. 3c). The spatial distribution peaks in front of the passages and the angular distribution sharply peaks at zero degree (Fig. 3d). In addition, we found that this directionality strengthens as the number of rows in the phononic crystal is increased and as the neck is reduced (Supplementary Information).

Thus, ballistic phonon transport can indeed occur in aligned lattices. Let us consider this process in more details. Ballistic transport suggests that phonons can travel, at least until a phonon-phonon scattering event, experiencing either no scattering at all or only *specular* surface reflections—as opposed to *diffuse* scattering reflections resulting in random scattering angles[35]. The probability ($p$) of specular scattering depends[36] on the phonon wavelength ($\lambda$), the root mean square surface roughness ($\eta$) and incidence angle ($\theta$) as $p = exp\ (-16\ \pi^2\ \eta^2\ cos^2\theta\ /\lambda^2)$. The wavelength depends on the temperature and shortens from ~25 nm at 4 K down to a few nanometres[37] at room temperature. Assuming $\lambda = 3$ nm and $\eta = 2$ nm (Methods), we can estimate that the scattering on hole surfaces can be partly specular ($p > 0$) even at room temperature if phonons approach the surfaces tangentially ($\theta > 80°$), as it happens in the passages between the holes of the aligned lattice[38,39].

Indeed, even at room temperature, similar directional heat fluxes in phononic crystals with aligned lattice have been shown via simulations based on the Boltzmann transport equation for phonons[39,40]. Tang *et al*[39] showed that the impact of specular surface reflections grows as the structure was scaled down. Assuming purely specular surface reflections ($p = 1$), they predicted a 15% difference between thermal conductivities of aligned and staggered structures, whereas some other theoretical works[38,41] captured no significant difference, assuming purely diffuse scattering ($p = 0$). Recently, Hao *et al*.[42] considered aligned lattices in both purely diffusive and purely specular scattering approximations; they found that the difference between thermal conductivities in these two cases strengthens as the neck becomes smaller and reaches 32% at the neck of 30 nm, as consistent with our results. Thus, our experimental data can be explained only by the presence of specular surface reflections and ballistic phonon transport in the passages between the holes.

To explain the temperature dependence of this effect, we should take into account that phonon wavelength becomes longer at lower temperatures, thus the range of incident angles, for which phonons can travel ballistically, becomes wider. Hence, the increased number of ballistic phonons. Similarly, the bulk mean free path—the average distance that phonons can travel until a phonon-phonon scattering event—also lengthens at lower temperatures[37,43,44], thus phonons can travel ballistically over longer distances.

## Coupling ballistic phonons into nanowires

The observed directionality of phonons in the aligned lattice implies that the phononic nanostructures can act as guides and directional sources of ballistic phonons. To demonstrate the directionality of phonons exiting the phononic crystals, we fabricated our second set of samples, in which phononic crystals with an aligned lattice ($a$ = 320 nm, $d / a$ = 0.84) and containing 10 rows of holes were connected to nanowires with a width of 120 nm and length in the 1 – 6 μm range, as shown in Figure 4a. These samples were of two types: *coupled* and *uncoupled*. In the coupled samples the nanowires were placed in front of the passages between the holes (Fig. 4b), in the maximums of the phonon spatial distribution discussed above. In this configuration phonons with low exit angles could ballistically pass from the phononic crystal directly into the nanowires. In the uncoupled samples, the nanowires were placed behind the holes (Fig. 4c), thus the direct passages into the nanowires were blocked.

The type of heat transport in nanowires is known to be length dependent[12,13,45], being mostly ballistic in nanowires of few micrometres in length, but becoming diffusive as the length is increased. This fact makes our experiment more challenging because we can expect not only a significant difference between decay times of coupled and uncoupled samples, but also a dependence on the nanowire length.

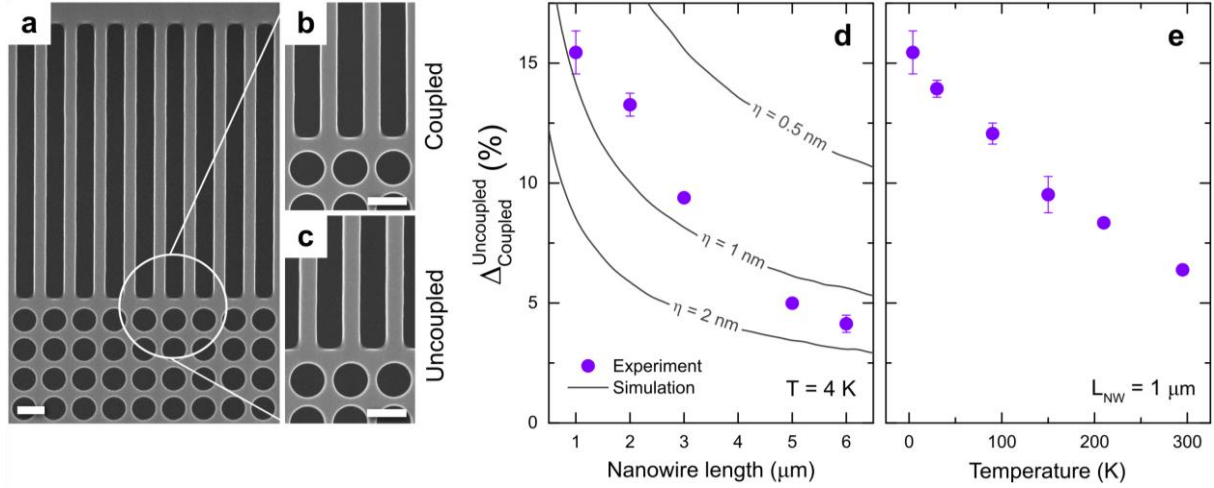

**Figure 4**. **Phonon coupling into nanowires**. **a-c**, SEM image of a typical nanowire-coupled phononic crystal sample with close-up views showing the difference between coupled and uncoupled samples. Scale bars are 300 nm. **d-e**, The experimentally measured difference between nanowire-coupled and uncoupled samples ($\Delta_{\text{Coupled}}^{\text{Uncoupled}} = (\tau_{\text{Uncoupled}} - \tau_{\text{Coupled}})/\tau_{\text{Coupled}}$) is length and temperature dependent, which shows that the phononic crystal acts as a directional source of ballistic phonons. Monte-Carlo simulations, represented by the lines in **d**, show that the relative difference is roughness dependent with a fair agreement for $\eta = 1$ nm.

Figure 4d shows the relative difference $\Delta_{\text{Coupled}}^{\text{Uncoupled}} = (\tau_{\text{Uncoupled}} - \tau_{\text{Coupled}})/\tau_{\text{Coupled}}$ between decay times measured on nanowire-coupled ($\tau_{\text{Coupled}}$) and uncoupled ($\tau_{\text{Uncoupled}}$) samples as a function of the nanowire length ($L_{\text{NW}}$) at 4 K. In the samples with short nanowires ($L_{\text{NW}} = 1$ μm), heat dissipates 15% faster through the coupled structure than through the uncoupled one. But, as the nanowires lengthen, this difference gradually decreases and almost disappears at the length of 6 μm. This experiment shows that the phononic crystal indeed creates a flux of phonons parallel to the nanowire axis. This flux enforces ballisticity in short nanowires, causing the significant difference between coupled and uncoupled samples. But this difference gradually disappears as the heat transport eventually becomes diffusive in long nanowires[12,45], regardless of the initial directionality.

To show that the faster heat dissipation in the coupled configuration comes from the directionality of phonons entering the nanowires, we simulated phonon transport in the nanowires with two distributions of initial angles: Lorentzian, representing coupled configuration, and uniform, representing uncoupled one. Measuring the average time that phonons stay in the nanowires, we can compare the simulation and experimental results. Figure 4d shows that the simulation results naturally depend on the surface roughness and match our experimental data for $\eta = 1$ nm, which is consistent with our estimations of surface roughness ($\eta \leq 2$ nm).

Moreover, the effect weakens as temperature is increased (Fig. 5c), which shows again that the effect is linked to wavelength and mean free path. Indeed, since bulk mean free path in silicon at room temperature is in the $0.1 - 10$ μm range[43,44,46], less than half of the phonons can traverse ballistically even the 1-μm-long nanowires[46]. However, these experiments show that the phononic crystal acts as a directional source of ballistic phonons, and therefore can be used for a variety of other applications.

## Heat focusing

One of the possible applications of these findings is heat localization. To show its practical realization, we propose a converging thermal lens consisting of circular rows of holes with various diameters, but identical necks, as shown in Figure 5a. In such a structure, our Monte-Carlo simulations predicts that the heat fluxes from the passages between the holes will converge at the focal point forming a 115-nm-wide hot-spot located 0.5 μm away from the structure (Fig. 5b). To demonstrate this phenomenon, we fabricated a set of samples with the same converging lens but different positions (δ) of a narrow slit through which heat can dissipate from the system, as shown in Figure 5a. Additionally, we fabricated two sets of reference samples: the first without any holes at all and the second with inverted (diverging) lenses (Fig. 5c).

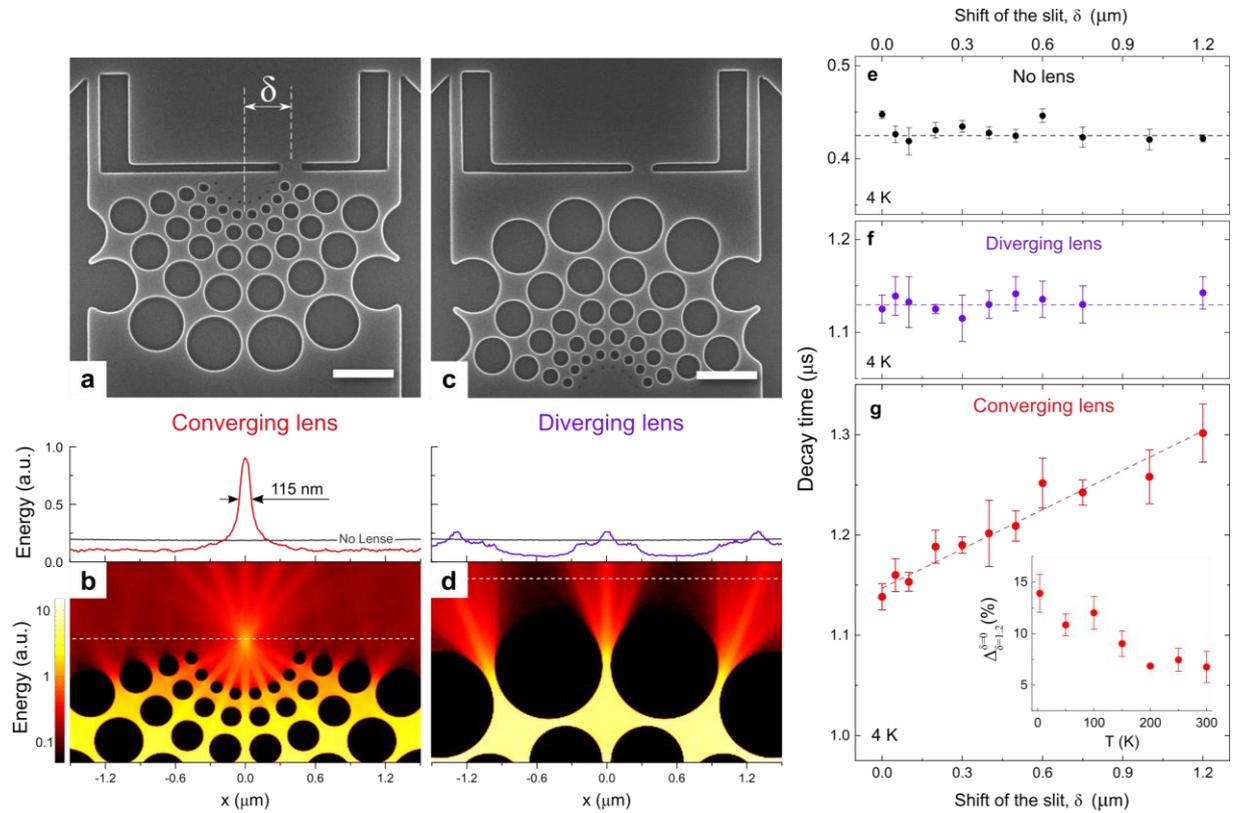

**Figure 5**. **Heat focusing**. SEM images of **a,** converging and **c,** diverging thermal lens samples with slits for heat dissipation. Scale bars are 1 μm. **b**, Monte-Carlo simulation predicts the formation of a hot-spot in the focal point of the converging lens, but, **d**, dispersion of heat in the diverging lens. **e**-**g**, Experimentally measured thermal decay times show no impact of the slit position in the samples without a lens or with the diverging lens, but strong dependence in the samples with the converging lens. The inset shows that the relative difference between decay times at δ = 0 and 1.2 μm decreases with temperature. Error bars show the standard deviation between different measurements on the same sample.

In the reference samples, heat is evenly dispersed in all directions, as shown by the simulation in Figure 5d. As a consequence, the measured thermal decay times are nearly independent of the slit position for both types of reference samples (Fig. 5e-f). In the converging lenses, however, heat dissipates faster when the slit is at the focal point (δ = 0 μm), but the dissipation slows down as the slit is moved aside (Fig. 5g). This result implies that the converging thermal lens indeed focuses heat in the focal point: the closer the slit, the faster heat can escape. The effect is also temperature dependent. The inset in Figure 5g shows that the difference ($\Delta_{\delta=1.2}^{\delta=0}$) between the decay times at different slit positions ($\delta = 0$ and 1.2 µm) decreases from 14% at 4 K to about 6% at room temperature, because phonon directionality weakens with temperature, as discussed above.

Another application of the thermal lens structure, very much sought after, is a tuneable thermal diode. Indeed, should the slits be placed on both sides of the lens, the heat propagation would depend on the slit position in the direction of converging lens, but not in the opposite direction. Hence the possibility of tuneable thermal rectification.

## Conclusions and outlook

In this work we showed that ray-like heat manipulation becomes possible at the nanoscale by using phononic crystals nanostructures. First, the comparison of aligned and staggered lattices at different scales and temperatures revealed the possibility of ballistic phonon transport in the aligned lattice. Next, the proof-of-concept experiments showed that these ballistic phonons can even be collected into nanowires, where their ballistic path is continued. These results imply that phononic crystals can guide and emit directional fluxes of ballistic phonons in solids. To show a potential for practical applications, we introduced a thermal lens nanostructures and showed evidence of nanoscale heat focusing. A more direct demonstration of the nanometre-scale hot-spot can be achieved using thermal mapping techniques[47], and should be the subject of a future work. However, we believe that this concept can not only compete with the conventional photothermal approach[3,48,49] to local heating, but also enhance it by placing the thermal lenses around the heating point. In addition, the thermal lens can be used to create a tuneable thermal diode, with the positions of the slits controlling the thermal rectification coefficient. One can imagine various other nanostructures that can guide, turn, confine or disperse the heat flux using different hole patterns. As we have shown, the efficiency of this approach depends on the dimensions and quality of the structures. Thus, as the advancements in nanofabrication technology enable further scaling down, complete control over directionality of the ballistic heat flux in nanostructures becomes a reality in the near future.

## Methods

**Sample fabrication.** All samples were fabricated on a commercially available silicon-on-insulator (SOI) wafer with a 1-µm-thick buried $SiO_2$ layer and a 145-nm-thick top monocrystalline (100) silicon layer. First, the electron-beam lithography was performed (with ZEP 520A as a resist) in order to deposit 125-nm-thick aluminium pads using electron beam physical vapour deposition (Ulvac EX-300), which was followed by the resist lift-off. Next, the structures were formed in the top silicon layer by the second electron-beam lithography followed by the reactive ion etching by means of an inductively coupled plasma system (Oxford Instruments Plasmalab System 100 ICP), with $SF_6/O_2$ gas as the etchant. Finally, the buried oxide layer was removed using vapour etching with diluted hydrofluoric acid. The hole diameters were measured via SEM for all samples, with inaccuracy of ± 2 nm.

This study is designed as comparative between pairs of similar samples, thus systematic errors are excluded from consideration, as most of the geometrical parameters (density of the holes, porosity, volume of material etc.) are exactly the same in both samples (aligned and staggered, coupled and uncoupled). Moreover, the samples of each set were fabricated simultaneously on the same wafer, thus we expect no variations in the surface roughness, as well as no difference in the rates of impurity and phonon-phonon scattering processes. For aligned and staggered lattices two identical sets of samples were fabricated and measured.

**Surface roughness considerations.** Figure 6a shows an atomic force microscope image of the top surface of the sample. The maximum surface irregularity does not exceed 0.5 nm, thus the root mean square of surface roughness ($\eta$) is well below this value and the roughness of the top and bottom surfaces is negligible. Figures 6b and 6c show SEM images of a neck between two holes and the side wall of a nanowire (top view), respectively. Although it is impossible to measure surface roughness by SEM accurately, we can estimate that the amplitude of the surface irregularity does not exceed 4 nm. Since the root mean square of the roughness is less than its maximum amplitude, we assume η ≤ 2 nm. A cross-sectional SEM image (Fig. 6d) also indicates low surface roughness and vertical hole profiles.

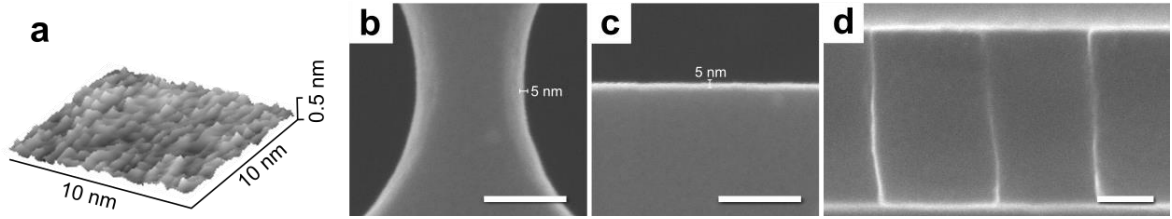

**Figure 6. Surface roughness. a**, AFM image of the top surface shows mean surface roughness of less than 0.5 nm. **b-c**, SEM images of a neck between two holes and the border of a nanowire show that surface irregularity is lower than 5 nm bar. **d**, Cross-sectional SEM image shows vertical hole profiles. Scale bars are 50 nm.

**Micro time-domain thermoreflectance.** The samples were placed in a He-flow cryostat mounted on motorized linear stages. This enabled us to perform measurements in vacuum (< $10^{-4}$ Pa) with precise control over the temperature (± 0.02 K) and position of the samples. A red light diode illuminated a large area on the surface of the wafer, making it possible to see the samples. A pulsed *pump* laser (λ = 642 nm, length of the pulse of 1 μs, repetition rate of 1 kHz) and a continuous-wave *probe* laser (λ = 785 nm) beams were focused on the aluminium pad on top of the studied sample via an optical (x40) objective. The reflection of the probe beam was continuously monitored by a silicon photodiode detector connected to a digital oscilloscope. Each 1-μs-long pulse of the pump beam caused a leap of the temperature in the aluminium pad (~ 300 nW is absorbed) followed by relaxation as heat dissipated through the sample (the average temperature increase in the structure does not exceed 5 K). This process was detected as a leap in the probe beam signal intensity caused by the change in reflectivity (ΔR), due to the heating, followed by a gradual return to the background value as the heat dissipated. A typical normalized ΔR/R signal as a function of time is shown in Figure 1b. The system recorded and averaged the ΔR/R decay functions over $10^4$ pulses of the pump beam, and automatically fitted the result by an exponential decay function exp(−*t* / τ), with *t* as the time and τ as a fitting parameter. To ensure the accuracy of the measurement, this iteration was repeated until the standard deviation of τ in the last 30 iterations was less than 1%, at which moment the value of τ was recorded.

To eliminate inaccuracies that might be caused by the laser beam alignment and sample fabrication, two identical samples of each type (only for aligned/staggered samples) were measured, each sample at least twice on different days. Thus each data point represents an average between several measurements. Error bars are calculated as the standard deviation, and do not exceed 5% for most of the samples.

**Monte-Carlo simulations.** The algorithm[35] traces trajectories of particle-like phonons (wave packets) in samples identical to those measured experimentally. The phonons are generated with random coordinates within a 4 × 4 µm$^2$ region at the centre of the system, representing the aluminium pad, and start moving in random initial directions; the angular frequencies (ω) are randomly assigned according to the Plank distribution at T = 5 K, calculated within the Debye approximation. When phonons encounter membrane or hole boundaries, the type of surface scattering (specular or diffuse) is determined by the specularity parameter (*p*), as discussed in the main text[36]. Since the surface roughness of the top layer is negligible, we assumed 100% specular scattering (*p* = 1) for the top and bottom surfaces, and performed simulations only in two dimensions. For other boundaries we assumed η = 2 nm, unless stated otherwise. To take into account internal scattering processes, at every moment a phonon can be scattered in random direction with a probability of 1 − *exp*( *t* / τ), where *t* is the time since the previous internal scattering event and τ is characteristic internal scattering time given[50] by $\tau^{-1} = \tau^{-1}_{impurity} + \tau^{-1}_{phonon-phonon}$, with $\tau^{-1}_{impurity} = 2.95 \times 10^{-45} \omega^4$ and $\tau^{-1}_{phonon-phonon} = (2\tau^{-1}_{TA} + \tau^{-1}_{TA})/3$, where $\tau^{-1}_{TA} = 9.3 \times 10^{-13} \omega T^4$ and $\tau^{-1}_{LA} = 2.0 \times 10^{-24} \omega^2 T^3$. The simulation time is set so as to reach the steady state when the input power is equal to the power exiting the system. Once the steady state is reached, the data are acquired during ten nanoseconds. To ensure good statistics, the number of phonons in the system is kept to several millions.

# References

1. Pop, E. Energy dissipation and transport in nanoscale devices. *Nano Res.* **3,** 147–169 (2010).
2. Kanatzidis, M. G. Nanostructured Thermoelectrics: The New Paradigm? *Chem. Mater.* **22,** 648–659 (2010).
3. Baffou, G. & Quidant, R. Thermo-plasmonics: Using metallic nanostructures as nano-sources of heat. *Laser Photonics Rev.* **7,** 171–187 (2013).
4. Roberts, N. A. & Walker, D. G. A review of thermal rectification observations and models in solid materials. *Int. J. Therm. Sci.* **50,** 648–662 (2011).
5. Li, N. *et al.* Phononics: Manipulating heat flow with electronic analogs and beyond. *Rev. Mod. Phys.* **84,** 1045–1066 (2012).
6. Wilson, R. B. & Cahill, D. G. Anisotropic failure of Fourier theory in time-domain thermoreflectance

experiments. *Nat. Commun.* **5,** 5075 (2014).

7. Siemens, M. E. *et al.* Quasi-ballistic thermal transport from nanoscale interfaces observed using ultrafast coherent soft X-ray beams. *Nat. Mater.* **9,** 26–30 (2010).

8. Hu, Y., Zeng, L., Minnich, A. J., Dresselhaus, M. S. & Chen, G. Spectral mapping of thermal conductivity through nanoscale ballistic transport. *Nat. Nanotechnol.* **10,** 701–706 (2015).

9. Lee, J., Lim, J. & Yang, P. Ballistic Phonon Transport in Holey Silicon. *Nano Lett.* **15,** 3273–3279 (2015).

10. Bae, M.-H. *et al.* Ballistic to diffusive crossover of heat flow in graphene ribbons. *Nat. Commun.* **4,** 1734 (2013).

11. Xu, X. *et al.* Length-dependent thermal conductivity in suspended single-layer graphene. *Nat. Commun.* **5,** 3689 (2014).

12. Hsiao, T.-K. *et al.* Observation of room-temperature ballistic thermal conduction persisting over 8.3 μm in SiGe nanowires. *Nat. Nanotechnol.* **8,** 534–8 (2013).

13. Hsiao, T. K. *et al.* Micron-scale ballistic thermal conduction and suppressed thermal conductivity in heterogeneously interfaced nanowires. *Phys. Rev. B* **91,** 035406 (2015).

14. Johnson, J. A. *et al.* Direct Measurement of Room-Temperature Nondiffusive Thermal Transport Over Micron Distances in a Silicon Membrane. *Phys. Rev. Lett.* **110,** 025901 (2013).

15. Hamad-Schifferli, K., Schwartz, J. J., Santos, A. T., Zhang, S. & Jacobson, J. M. Remote electronic control of DNA hybridization through inductive coupling to an attached metal nanocrystal antenna. *Nature* **415,** 152–5 (2002).

16. Ichikawa, M., Ichikawa, H., Yoshikawa, K. & Kimura, Y. Extension of a DNA molecule by local heating with a laser. *Phys. Rev. Lett.* **99,** 148104 (2007).

17. Skirtach, A. G. *et al.* The role of metal nanoparticles in remote release of encapsulated materials. *Nano Lett.* **5,** 1371–1377 (2005).

18. Riehemann, K. *et al.* Nanomedicine - Challenge and perspectives. *Angew. Chemie - Int. Ed.* **48,** 872–897 (2009).

19. Lee, K.-D. *et al.* Thermoelectric Signal Enhancement by Reconciling the Spin Seebeck and Anomalous Nernst Effects in Ferromagnet/Non-magnet Multilayers. *Sci. Rep.* **5,** 10249 (2015).

20. Wuttig, M. & Yamada, N. Phase-change materials for rewriteable data storage. *Nat. Mater.* **6,** 824–832 (2007).

21. Hamann, H. F., O'Boyle, M., Martin, Y. C., Rooks, M. & Wickramasinghe, H. K. Ultra-high-density phase-change storage and memory. *Nat. Mater.* **5,** 383–387 (2006).

22. Jin, C. Y., Li, Z., Williams, R. S., Lee, K. C. & Park, I. Localized temperature and chemical reaction control in nanoscale space by nanowire array. *Nano Lett.* **11,** 4818–4825 (2011).

23. Maldovan, M. Sound and heat revolutions in phononics. *Nature* **503,** 209–217 (2013).

24. Marconnet, A. M., Asheghi, M. & Goodson, K. E. From the Casimir Limit to Phononic Crystals: 20 Years

of Phonon Transport Studies Using Silicon-on-Insulator Technology. *J. Heat Transfer* **135,** 061601 (2013).

25. Maldovan, M. Phonon wave interference and thermal bandgap materials. *Nat. Mater.* **14,** 667–674 (2015).

26. Ravichandran, J. *et al.* Crossover from incoherent to coherent phonon scattering in epitaxial oxide superlattices. *Nat. Mater.* **13,** 168–72 (2014).

27. Jain, A., Yu, Y.-J. & McGaughey, A. J. H. Phonon transport in periodic silicon nanoporous films with feature sizes greater than 100 nm. *Phys. Rev. B* **87,** 195301 (2013).

28. Wagner, M. R. *et al.* Two-Dimensional Phononic Crystals: Disorder Matters. *Nano Lett.* **16,** 5661–5668 (2016).

29. Anufriev, R., Maire, J. & Nomura, M. Reduction of thermal conductivity by surface scattering of phonons in periodic silicon nanostructures. *Phys. Rev. B* **93,** 045411 (2016).

30. Nomura, M. *et al.* Impeded thermal transport in Si multiscale hierarchical architectures with phononic crystal nanostructures. *Phys. Rev. B* **91,** 205422 (2015).

31. Song, D. & Chen, G. Thermal conductivity of periodic microporous silicon films. *Appl. Phys. Lett.* **84,** 687–689 (2004).

32. Romano, G. & Grossman, J. C. Heat Conduction in Nanostructured Materials Predicted by Phonon Bulk Mean Free Path Distribution. *J. Heat Transfer* **137,** 071302 (2015).

33. Lim, J. *et al.* Simultaneous Thermoelectric Property Measurement and Incoherent Phonon Transport in Holey Silicon. *ACS Nano* **10,** 124–132 (2016).

34. Schierning, G. Silicon nanostructures for thermoelectric devices: A review of the current state of the art. *Phys. Status Solidi A* **211,** 1235–1249 (2014).

35. Ramiere, A., Volz, S. & Amrit, J. Geometrical tuning of thermal phonon spectrum in nanoribbons. *J. Phys. D. Appl. Phys.* **49,** 115306 (2016).

36. Soffer, S. B. Statistical Model for the Size Effect in Electrical Conduction. *J. Appl. Phys.* **38,** 1710 (1967).

37. Minnich, A. J. *et al.* Thermal Conductivity Spectroscopy Technique to Measure Phonon Mean Free Paths. *Phys. Rev. Lett.* **107,** 095901 (2011).

38. Fu, B., Tang, G. H. & Bi, C. Thermal conductivity in nanostructured materials and analysis of local angle between heat fluxes. *J. Appl. Phys.* **116,** 124310 (2014).

39. Tang, G. H., Bi, C. & Fu, B. Thermal conduction in nano-porous silicon thin film. *J. Appl. Phys.* **114,** 184302 (2013).

40. Romano, G. & Grossman, J. C. Toward phonon-boundary engineering in nanoporous materials. *Appl. Phys. Lett.* **105,** 033116 (2014).

41. Jean, V., Fumeron, S., Termentzidis, K., Tutashkonko, S. & Lacroix, D. Monte Carlo simulations of phonon transport in nanoporous silicon and germanium. *J. Appl. Phys.* **115,** 648–655 (2014).

42. Hao, Q., Xiao, Y. & Zhao, H. Characteristic length of phonon transport within periodic nanoporous thin

films and two-dimensional materials. *J. Appl. Phys.* **120,** 065101 (2016).

43. Regner, K. T. *et al.* Broadband phonon mean free path contributions to thermal conductivity measured using frequency domain thermoreflectance. *Nat. Commun.* **4,** 1640 (2013).

44. Zeng, L. *et al.* Measuring Phonon Mean Free Path Distributions by Probing Quasiballistic Phonon Transport in Grating Nanostructures. *Sci. Rep.* **5,** 17131 (2015).

45. Zhang, H., Hua, C., Ding, D. & Minnich, A. J. Length dependent thermal conductivity measurements yield phonon mean free path spectra in nanostructures. *Sci. Rep.* **5,** 9121 (2015).

46. Jiang, P., Lindsay, L. & Koh, Y. K. The role of low-energy phonons with mean-free-paths >0.8 um in heat conduction in silicon. *J. Appl. Phys.* **119,** 245705 (2015).

47. Reparaz, J. S. *et al.* A novel contactless technique for thermal conductivity determination: Two-laser Raman thermometry. *Rev. Sci. Instrum.* **85,** 034901 (2014).

48. Yannopapas, V. Localized heating of nanostructures by coherent laser pulses. *J. Phys. Chem. C* **117,** 14183–14188 (2013).

49. Govorov, A. O. & Richardson, H. H. Generating heat with metal nanoparticles. *Rev. Lit. Arts Am.* **2,** 30–38 (2007).

50. Randrianalisoa, J. & Baillis, D. Monte Carlo Simulation of Steady-State Microscale Phonon Heat Transport. *J. Heat Transfer* **130,** 072404 (2008).

**Author contributions**

Roman Anufriev designed the study, fabricated the samples, conducted the measurements, processed and interpreted the experimental data, and wrote the article. Aymeric Ramiere developed the Monte-Carlo model, performed numerical simulations and contributed to designing of the samples. Jeremie Maire fabricated the experimental setup, contributed to the sample fabrication and experimental data processing, and made a major effort in preparation of Supplementary Information. Masahiro Nomura contributed to the design of the study and development of the experimental setup, and supervised the work.

**Acknowledgements**

This work was supported by the Project for Developing Innovation Systems of the MEXT, Japan, Kakenhi (15H05869 and 15K13270), PRESTO JST and Postdoctoral Fellowship of Japan Society for the Promotion of Science. We also acknowledge Anthony George for assistance with AFM measurements and Ryoto Yanagisawa for assistance with sample fabrication.